%% file: prc_paper_auth.tex
\documentclass[prc,i10pt,twocolumn,english,aps,superscriptaddress]{revtex4-1}
\usepackage[T1]{fontenc}
\usepackage[latin9]{inputenc}
\usepackage{babel}
\usepackage{array}
\usepackage{graphicx}

\begin{document}

\title{Double $K^0_S$ Photoproduction off the Proton at CLAS}

%\author{S. Chandavar$^1$, K. Hicks$^1$, J. Goetz$^1$, M.C. Kunkel$^2$, 
%M. Paolone$^3$, D. Weygand$^4$ }
%\affiliation{$^1$Ohio University, $^2$Julich Institut fur Kernphysics
%$^3$ Temple University, $^4$ Jefferson Lab }

\input{auth_revtex.tex}

\date{\today}

\begin{abstract}
The $f_{0}(1500)$ meson resonance is one of several contenders to have
significant mixing with the lightest glueball.
This resonance is well established from several previous experiments.
Here we present the first photoproduction data for the $f_0(1500)$ 
via decay into the $K_{S}^{0}K_{S}^{0}$ channel using the CLAS detector. 
The reaction $\gamma p\to f_{J}p\rightarrow K_{S}^{0}K_{S}^{0}p$, 
where $J=0,2$, was measured with photon energies from 2.7 to 5.1 GeV. 
A clear peak is seen at 1500 MeV in the background subtracted invariant 
mass spectra of the two kaons.
This is enhanced if the measured 4-momentum transfer to the proton target 
is restricted to be less than 1.0 GeV$^{2}$.  
By comparing data with simulations, it can be concluded that the peak 
at 1500 MeV is produced primarily at low $t$, which is consistent 
with a $t$-channel production mechanism. 
\end{abstract}

\pacs{102454}

\maketitle

\section{Introduction }

The search for glueballs has been ongoing for several decades \cite{key-1}.
The lightest glueball has been predicted by quenched lattice QCD to
have a mass in the range of $1.5-1.8$ GeV and $J^{PC}=0^{++}$ \cite{key-2}.
The mixing of glueball states with neighbouring meson states complicates
their identification and hence possible glueball candidates have been 
extensively scrutinized. 

Of the scalar mesons, the isoscalars are the mesons of interest in
the search for glueballs. Five isoscalar scalars have been identified
by experiment and listed by the Particle Data Group (PDG): $f_{0}(500)$,
$f_{0}(980)$, $f_{0}(1370),$ $f_{0}(1500)$ and $f_{0}(1710)$ \cite{key-3}.
However, of these, only two can belong to the meson scalar nonet
(see the tentative assignments given in Ref. \cite{key-3}).
As discussed below, two of these states (the $f_0(500)$ and $f_0(980)$), 
are thought to be either meson-meson molecules or $qq\bar{q}\bar{q}$ 
states, but this still leaves three possible scalar mesons to fit 
into two quark-model slots.
The excess of scalar states suggests the presence of a glueball state,
with the same quantum number ($J^{PC}=0^{++}$),
which mixes with the scalar meson states \cite{key-1}.
By analyzing the decay channels and production mechanisms of these 
three scalar meson candidates, the glueball mixing can be compared 
with theoretical predictions.  

In reality, there is no consensus on the status of several of these 
scalars \cite{key-3}. For some scalar mesons, such as the $f_0(500)$, 
the distinction between resonance and background is 
difficult because of the large decay widths. 
Also, the opening of multiple decay channels within short
mass intervals makes the background shapes difficult to model
\cite{key-3}.  Yet the high interest for a possible glueball state 
(and how it mixes with the scalar mesons) motivates further 
measurements of the production mechanisms and decays of the scalars.

After many years and many experiments focused on the scalar mesons,
there is still confusion on how to classify these states \cite{key-3}. 
The $f_{0}(980)$ and the $a_{0}(980)$, along with the $f_{0}(500)$ and 
$K_{0}^{*}(800)$, likely form a low-mass nonet of primarily
four-quark states \cite{key-4, key-5}. 
Models based on unitary quarks with coupled $q\bar{q}$ and meson-meson 
channels interpret the scalars as two nonets, 
the \{$f_{0}(980)$, $a_{0}(980)$, $f_{0}(500)$ and $K_{0}^{*}(800)$\}, 
and the \{$f_{0}(?)$, $a_{0}(1450)$ and $K_{0}^{*}(1430)$\}, where 
$f_0(?)$ stands for two of the $f_0(1370)$, $f_0(1500)$ or $f_0(1710)$. 
These are two expressions of the same bare states \cite{key-3,key-4}, 
where the former nonet is consistent with a dominant 
$qq\bar{q}\bar{q}$ component.
In the latter nonet, the $f_{0}(1500)$ and the $f_{0}(1710)$ are
candidates for having the highest glueball content \cite{key-1}. 

Recently, there has been a resurgence of interest in the $f_0(1710)$ 
as the best glueball candidate based on Holographic QCD calculations 
\cite{key-6}.  In that paper, they calculate the glueball decay rates 
and find a suppressed decay of the glueball into final states with two 
pions (and also a very small coupling to four-pion decay). 
The decay ratio of $\Gamma(\pi\pi)/\Gamma(KK)$ for the $f_0(1710)$ 
is found \cite{key-3} to be much smaller than the $SU(3)_F$ value 
of 3/4, giving better agreement with their predictions. 
However, as pointed out above, the experimental measurements of 
these scalar meson decays is sometimes conflicting \cite{key-3}, 
and hence more measurements are needed.

\begin{figure}
\vspace{0.02\textwidth}
\begin{centering}
\includegraphics[scale=0.6]{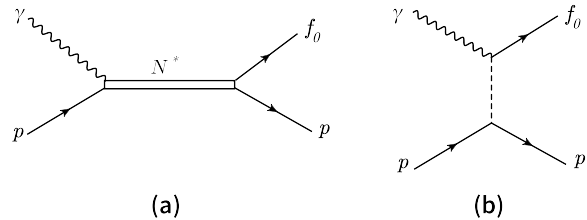}
\par\end{centering}
\protect\caption{
Schematic diagrams of reaction mechanisms for (a) $s$-channel and 
(b) $t$-channel photoproduction of a scalar meson.
\label{fig:s-and-t-channel}}
\end{figure}

Photoproduction has been suggested as a means to look for glueballs
\cite{key-7}. Production can occur primarily via two channels, as shown in
Fig. \ref{fig:s-and-t-channel}. In the $s$-channel, the
photon and proton interact to form an intermediate particle that 
then decays into a meson and a proton. This channel can couple directly 
to a scalar meson with high glueball content.
In the $t$-channel on the other hand, the photon must couple to 
the exchange particle. In this case, the outgoing particle (and 
hence the exchange particle) has neutral charge, and the photon 
coupling is suppressed.  For a pure $SU(3)_F$ glueball \cite{key-2}, 
made entirely of gluons (with no quark-antiquark pairs), there is 
no charge and hence no coupling to the photon.
For a meson with a large glueball admixture, the photon coupling 
in $t$-channel is expected to be partly suppressed \cite{brodsky}, 
since its wavefunction contains a glueball component. 

The $t$-channel strength can be separated, to a large extent, from 
$s$-channel by measuring the 4-vectors of the detected particles and 
calculating the momentum transfer, $|t|$, to the proton.  
Low values of momentum transfer typically correspond to $t$-channel diagrams, 
whereas $s$-channel diagrams span a wider range of momentum transfer.

Here, we examine the photoproduction of scalar and tensor mesons
at energies from $\sqrt{s}=$2.4 to 3.3 GeV, spanning an 
energy region above threshold to produce scalar $f_0$ mesons off a 
proton target. 
The following sections provide the experimental details and the 
analysis procedures used to study the $t$-dependence of the yeild 
for one of these states with a mass near 1500 MeV. 
While the statistics are low, making it difficult to draw firm 
conclusions on the spin $J$ of the peak at 1500 MeV, the data 
validates the technique, and future measurements with higher 
statistics at Jefferson Lab will provide more conclusive results.

\section{Experimental Setup}

The experiment was carried out in Hall B at the Thomas Jefferson National 
Accelerator Facility using the CEBAF Large Acceptance Spectrometer (CLAS) 
 \cite{key-8}.  
The primary electron beam from the CEBAF accelerator struck a 
gold foil of $10^{-4}$ radiation lengths, producing a tagged 
real photon beam \cite{key-9}.  
The photon energy was determined from the trajectory of the detected 
electron in the tagger focal plane. 
The initial electron energy for this experiment, called g12, was 
5.71 GeV and the tagged photon energy range was between 20\% to 95\% of the 
initial electron energy.   
The photon energy resolution depends on energy and was $<7.6$ MeV. 
The g12 data were taken from April to June, 2008, with a beam of 
polarized electrons (the photon beam polarization was not used in the 
present analysis).

The photons struck a liquid hydrogen target of length 40 cm and 
diameter 4 cm.  The target was placed 90 cm upstream of the center 
of CLAS in order to improve the acceptance for particles produced 
at small angles. Final state hadrons from the photon-nucleon 
interactions went into a toroidal magnetic field produced by the 
six-sector coils of the CLAS detector \cite{key-8}.  The coils 
were run with a current of 1930 A, which is half of the maximum 
design current. Positively charged particles were bent away from 
the beamline, thus having a larger detector acceptance than 
negatively charged particles of the same momentum.  

Particles were tracked using a set of three drift chambers in each 
sector \cite{key-10}, giving a momentum resolution of 
$\sim$0.5\% for charged particles of momentum $p=1$ GeV/c. The time of 
flight of the particles was measured between a start counter 
that surrounded the target \cite{key-11} and an array of 
scintillator bars that covered the exterior of the CLAS detector 
\cite{key-8}.  A photon in the tagger along with at least two charged 
particles in a timing coincidence produced a trigger for the 
data acquisition system.  Details of the trigger can be found 
in Ref.~\cite{key-12}.

\section{Analysis Procedures}

The reactions

\begin{equation}
\gamma p \rightarrow f_0 p \ \ {\rm and}\ \ 
\gamma p \rightarrow f_2 p
\label{eq:gammap to f0}
\end{equation}
were studied in the decay branch 
\begin{equation}
f_J \rightarrow K_S^0 + K_S^0
        \rightarrow \pi^+ \pi^- \pi^+ \pi^- \ .
\label{eq:fo to ksks}
\end{equation}
In the above reactions, the photon beam and the proton target interact
to produce a scalar (tensor) meson and the proton. The scalar (tensor)
meson then decays into a pair of short lived neutral kaons ($K_{S}^{0}$),
each of which decay into a pair of charged pions. The final state
particles are $\pi^+\pi^-\pi^+\pi^- p$, of which the four charged pions
are detected, while the proton is identified via the missing mass
technique. Requiring the final state to be $K_{S}K_{S}$ (four
detected pions) ensures that the $CP$ of the resonant meson is $++$.
This limits the final state meson to have even $J$, and we expect
$J=0,2$ to dominate near threshold.

The trigger configurations, calibrations of the detector sub-systems, and 
determination of the photon flux have been detailed in Ref. \cite{key-12}.

\subsection{The Basic Cuts}

The basic analysis cuts (event selection criteria) and momentum corrections 
that are applied to the data are listed in Table \ref{tab:The-basic-cuts}.
These will be discussed in Sections \ref{sub:Timing-Cut} through
\ref{sub:Sideband-Subtraction}. Kinematic cuts are described 
in Section \ref{sub:t-cuts}.

\begin{table*}
\begin{tabular}{|c|c|c|}
\hline 
Cut Level & Type of Cut & Size of Cut\\
\hline 
1 & Timing Cut for identification of pions & $\pm1$ ns\\
\hline 
2 & Fiducial Cut & Fit to CLAS acceptance\\
\hline 
3 & Missing mass (proton)  & $\pm$0.0497 GeV (3$\sigma$)\\
\hline 
4 & Photon beam energy  &  2.7-3.0 and 3.1-5.1 GeV\\
\hline 
5 & $K_{S}^{0}$ peak and sideband subtraction & 0.01614 GeV (3$\sigma$)\\
\hline 
\end{tabular}

\protect\caption{The event selection criteria (cuts) used in this analysis. 
\label{tab:The-basic-cuts}}
\end{table*}

\subsubsection{Timing Cut\label{sub:Timing-Cut}}

During the time that the DAQ recorded one event, several photons could
be measured by the tagger. Of these photons, it was necessary
to find that photon which interacted with the target to produce the
particles in CLAS. The tracks measured in the drift chamber (DC) were
extrapolated to the start counter and also to the Time-of-Flight (TOF)
scintillator bars. Using time and distance measurements, the start time for
every track was calculated. The beam RF time corresponding to the 
start times for all tracks, corrected for the vertex position 
in the target, was taken as the event vertex time.

To identify and select the detected particles as pions, the TOF Difference
method was employed. In this method, the difference between the calculated
and measured time of flight was constrained to be within 1 ns. The
calculated TOF was determined in the following manner: the mass of
the particles was assumed to be the mass of the charged pion,
139.57 MeV. Then, using the measured momentum of the particle, we
can calculated the time required by the $\pi^{+}$or $\pi^{-}$ to
traverse the path, $L_{sc}$, from the target to the TOF:
\begin{equation}
\beta_{calc}=\frac{p_{measured}}{\sqrt{p_{measured}^{2}+m_{\pi}^{2}}}
\end{equation}
and
\begin{equation}
TOF_{calc}=\frac{L_{sc}}{c\beta_{calc}} \ ,
\end{equation}
where $c$ is the speed of light.

The measured TOF is the difference in time of the scintillator hit, 
$t_{sc}$, and the event vertex time,
\begin{equation}
TOF_{measured}=t_{sc}-t_{vertex} \ .
\end{equation}
The difference between the measured and calculated TOF, 
\begin{equation}
\Delta TOF=TOF_{measured}-TOF_{calc} \ ,
\end{equation}
was calculated and a $\pm 1.0$ ns cut on $\Delta TOF$ was applied. 
If this cut led to the selection of at least two positively charged 
pions and at least two negatively charged pions, then the event was 
passed on for further analysis. 

The photon whose vertex time matched most closely to the average start
counter time of the pions was chosen. Depending on the electron beam current,
there could be more than one ``good'' photon. Using the four-momenta of
the four pions, the target proton and the photon, the missing mass off of
the four pions was calculated.  If a single photon was within the 
missing mass cut (see Cut 3 of Table \ref{tab:The-basic-cuts}) then 
this photon was chosen. After this selection,
the events with one good photon accounted for 96\% of the total
events with 2$\pi^{+}$ and 2$\pi^{-}$.

\subsubsection{Fiducial Cut}

The CLAS torus magnet consisted of six superconducting coils arranged 
to form a toroid around the beamline. In the rare case where one of 
the decay particles hit support material and scattered into the detector, 
an improper track would be observed. 
Also particle tracks reconstructed very near the 
coils could be inaccurate due to slight distortion of the magnetic field. 
Therefore, it is useful to apply fiducial cuts to reject those particles 
that track into the regions immediately surrounding the coils. 
Such cuts were employed here, which trimmed a few percent off the 
edges of the active region of the CLAS detector.  Details on the 
fiducial cuts are available elsewhere \cite{key-12}.

\subsubsection{Energy Loss Corrections}

To account for the energy loss of the decay particles while traversing
through the target, start counter and their associated assembly materials,
the CLAS $eloss$ package \cite{key-14} was employed.
It corrected for the loss of energy using the Bethe-Bloch equation,
which relates the energy loss of a particle through a material with
the characteristics of the material and the distance traveled by
the particle in that material. 
This software package had all of the geometry of the target and the 
surrounding material, so that each track was corrected for energy loss 
according to its trajectory.

\begin{figure}
\begin{centering}
\includegraphics[scale=0.45]{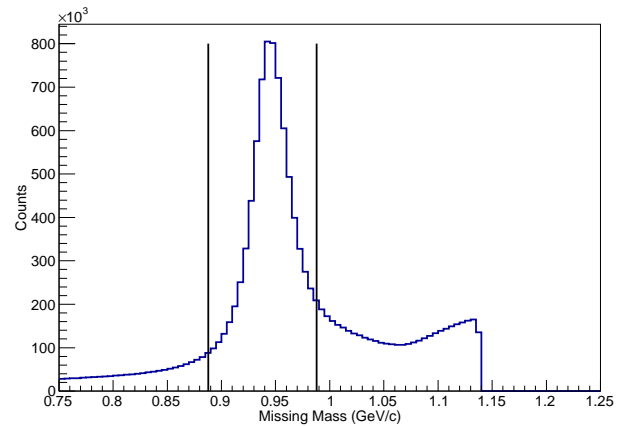}
\par\end{centering}
\protect\caption{
Missing mass for the reaction $\gamma p \to 2\pi^+ 2\pi^- X$ was calculated 
from the incident photon energy, the target proton mass and the momentum 
of the detected charged pions. 
A clear peak is seen at the mass of the proton. 
The vertical lines enclose the selected events.
\label{fig:missing-mass} }
\end{figure}

\subsubsection{Missing Mass Cut}

The missing particle in the reaction is calculated using the four-momenta
of detected pions, beam and target: 

\begin{equation}
P_{miss} =(P_{beam} + P_{target}) - 
(P_{\pi^+} + P_{\pi^-} + P_{\pi^+} + P_{\pi^-}) \ .
\end{equation}
The missing particle was then defined to be the proton by selecting
those events that had a missing mass within $\sim$50 MeV of the mass 
of the proton (Cut 3 in Table \ref{tab:The-basic-cuts}), as shown 
in Fig. \ref{fig:missing-mass}. Only the particle identification 
and fiducial cuts are applied in Fig. \ref{fig:missing-mass}.  
The small background under the proton peak was significantly reduced 
after further analysis cuts were employed.

\subsubsection{Beam Energy Cut}

The threshold photon energy for the production of the $f_{0}(1500)$, 
which is the particle of main interest, can be calculated by means
of the following equation:
\begin{equation}
E_{\gamma^{min}}=\frac{m_{f_{0}(1500)}^{2}
+2m_{f_{0}(1500)}m_{p}}{2m_{p}} \ .
\label{eq:Threshold}
\end{equation}
From this, the minimum energy to produce a $f_{0}(1500)$
in this reaction is $E_{\gamma^{min}} = 2.7$ GeV. 
Since we are interested in studying the $f_0(1500)$, photon 
energies below 2.7 GeV were removed in further analysis. 
For the g12 experiment, there is a discontinuity in $E_{\gamma}$
at $\sim3$ GeV due to a bad timing counter in the photon tagger.
This region is excluded from the analysis by eliminating the events
between 3.0 and 3.1 GeV for both data and simulations. 
This event selection (Cut 4 in Table \ref{tab:The-basic-cuts}) has 
been applied to all of the following figures.

\begin{figure}
\begin{centering}
\includegraphics[scale=0.45]{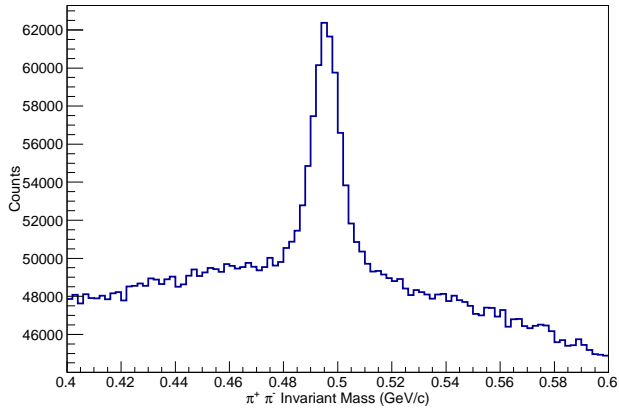}
\par\end{centering}
\protect\caption{
Invariant mass spectrum for one combination of $\pi^+\pi^-$. 
A clear $K_S^0$ peak is seen over a two-pion background. 
Other combinations show a similar mass distribution.
\label{fig:pi-pi Invariant mass} }
\end{figure}

\subsubsection{Sideband Subtraction\label{sub:Sideband-Subtraction}}

The four pions, $\pi_{1}^{+},\pi_{1}^{-},\pi_{2}^{+}$ and $\pi_{2}^{-}$
can form $2K_{S}^{0}$ in two ways. We use the following naming convention:
\begin{equation}
K1 = \pi_1^+ \pi_1^- , \ \ \  K2 = \pi_2^+ \pi_2^- 
\end{equation}
and
\begin{equation}
K3 = \pi_1^+ \pi_2^- , \ \ \  K4 = \pi_2^+ \pi_1^- \ .
\end{equation}
The numbering of the pions was based on the order in which they were
recorded by the event builder software. In order to avoid any bias, 
the ordering was randomized in our analysis.
In a given event, the 4 pions can form either: (a) $K1$ and $K2$
or (b) $K3$ and $K4$. Figure \ref{fig:pi-pi Invariant mass} shows
the invariant mass spectrum for the first pair of $\pi^{+}\pi^{-}$,
which shows a clear $K_{S}^0$ peak above a nearly flat background. 
Because we randomized the ordering of the pions, other combinations 
show similar invariant mass distributions.  

If the invariant masses of the pairs of pions are plotted against
one another, a strong correlation is observed (if one pair forms a 
$K_S^0$, then the other pair is very likely to form a second $K_S^0$,
indicating a common decay source for a majority of the events). 
In order to reduce background under the $K_S^0$ peak, a standard
method of background subtraction is used. A $3\sigma$ region
(see Cut 5 of Table \ref{tab:The-basic-cuts}) is applied around 
the $K_S^0$ mass to identify events lying in the signal region. 
Since the background is relatively flat, the bands on either side 
of the signal can be considered to be the average background below 
the $K_{S}$ peak. These regions are of the same width as the signal 
region and are referred to as sidebands.

\begin{figure*}
\begin{centering}
\includegraphics[scale=0.50]{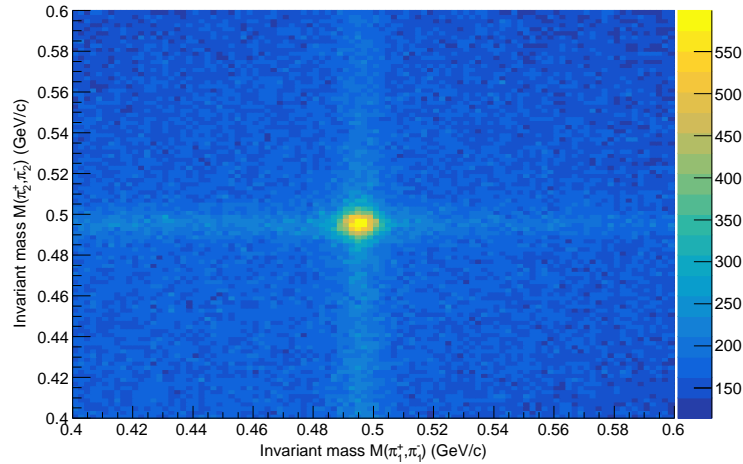}
\par\end{centering}
\protect\caption{
Correlation between the invariant mass of one $\pi^+ \pi^-$ pair 
and the other $\pi^+ \pi^-$ pair.  The spot at the center shows 
the situation where both pion pairs come from decays of two $K_S^0$. 
\label{fig:2D sideband} }
\end{figure*}

A 2-dimensional plot of the invariant masses of one pair
of pions versus the other is shown in Fig. \ref{fig:2D sideband}, 
where a clear $K_S^0$-$K_S^0$ correlation is seen. 
The signal region is a square centered on the $K_S^0$ mass 
(on each axis).
Also, there are several sideband regions to consider. 
Each sideband region is a square, sharing one edge (or one corner) 
with the signal region and with its center offset by $6\sigma$ from 
the center of the signal region.
Note that there are faint horizontal and vertical lines that
go through the signal and sideband regions.  These are likely 
due to events where one $K_S^0$ and a strange baryon resonance 
($\Sigma^*$) were produced, followed by a decay such as 
$\Sigma^{*+} \to \Lambda \pi^+ \to p \pi^- \pi^+$.
These events were subtracted, in the correct proportion, from 
the background under the signal region by using the sidebands.

\begin{figure*}
\begin{centering}
\includegraphics[scale=0.40]{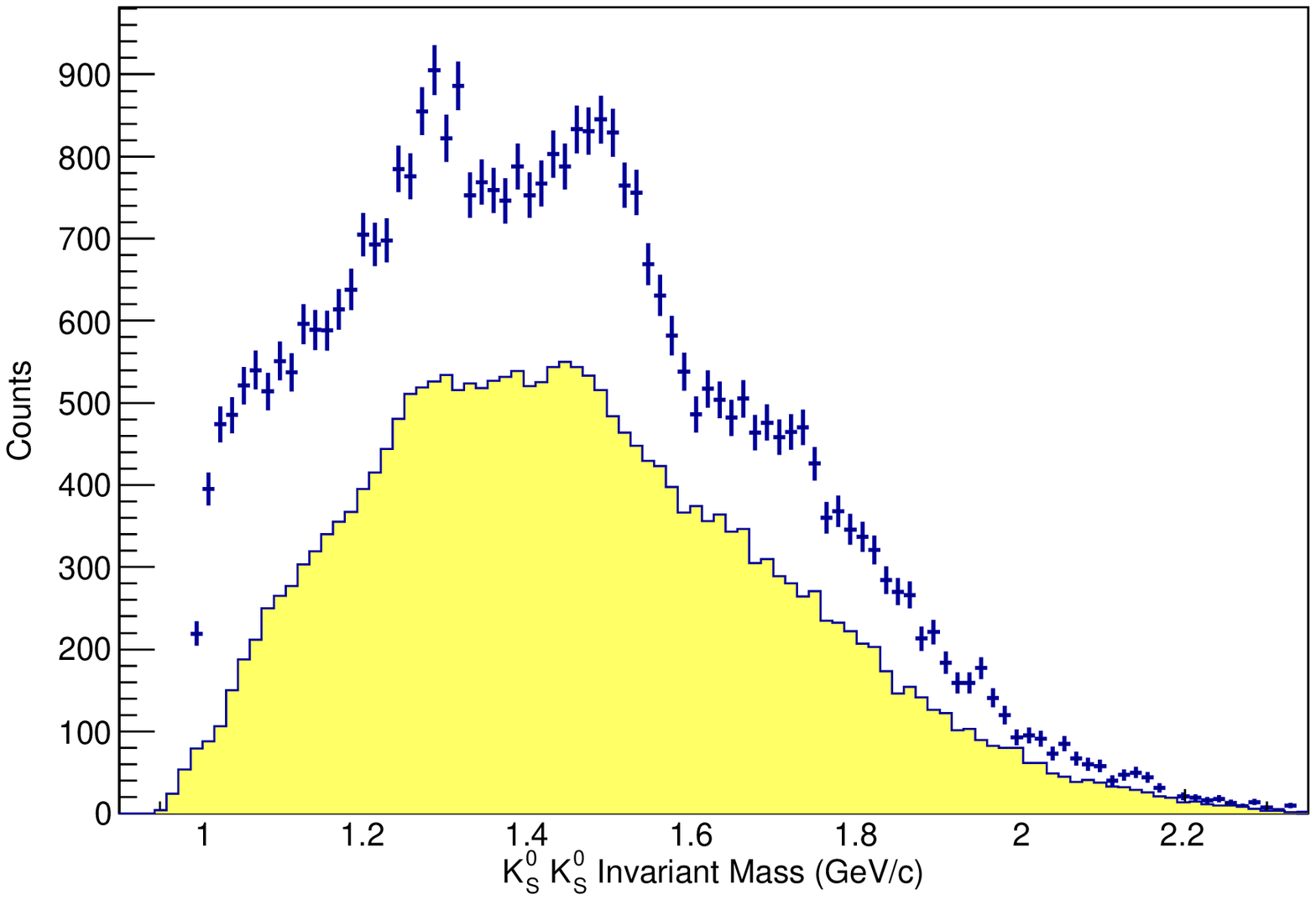}
\includegraphics[scale=0.40]{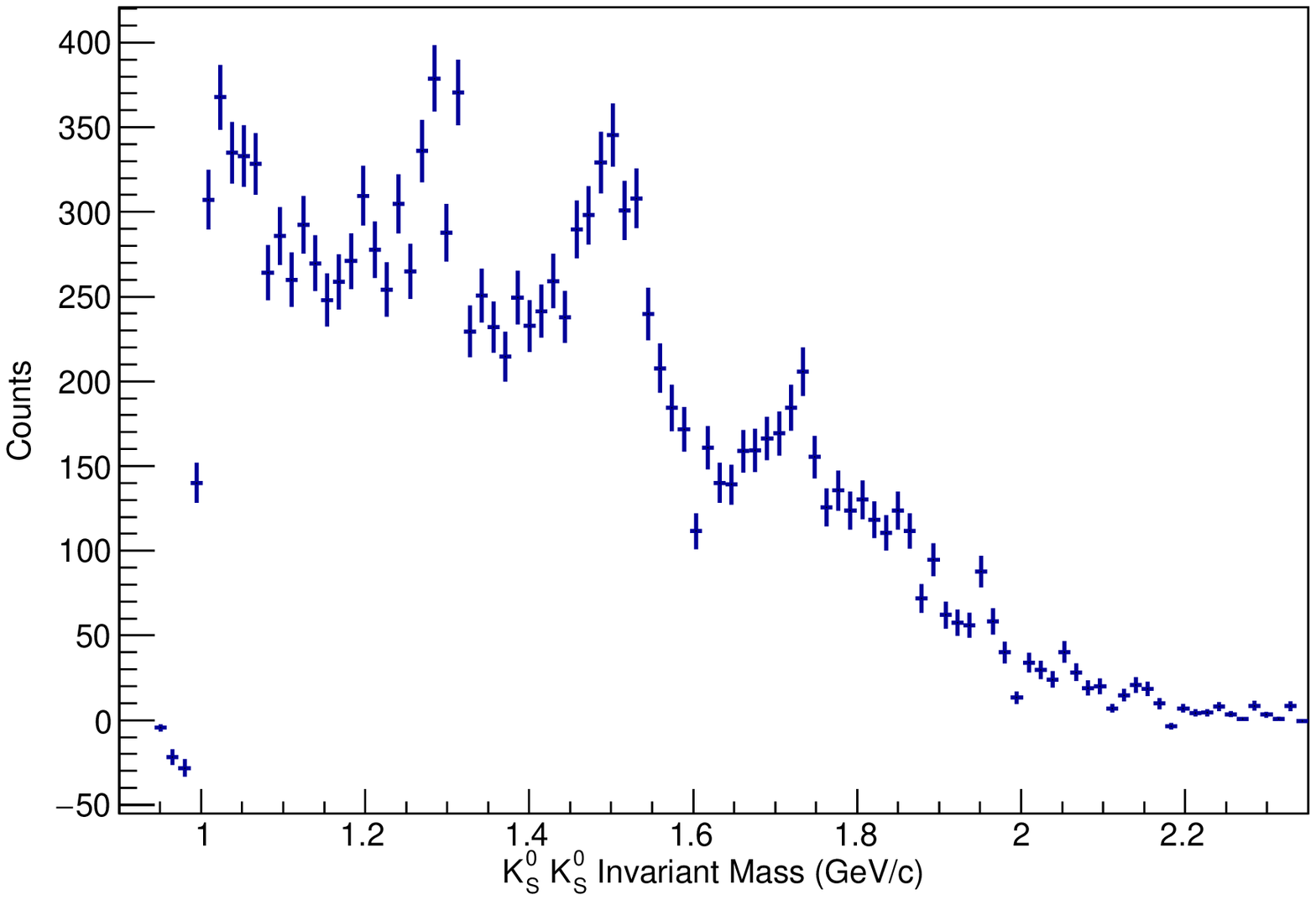}
\par\end{centering}
\protect\caption{
The left histogram with the error bars shows the signal + background,
whereas the shaded (yellow online) histogram is the average 
sideband background. 
The right histogram is the sideband subtracted histogram.  
The peak near 1.50 GeV is the region of interest. 
\label{fig:Mkkbs} }
\end{figure*}

Figure \ref{fig:Mkkbs} shows the $4\pi$ invariant mass spectrum before
and after background subtraction.  This mass spectrum is virtually 
identical to that formed from the $2K_S^0$ invariant mass. 
We choose to plot it this way because the average of the sideband 
regions are plotted together with the signal region.
Two clear peaks, one at $\sim 1.28$ GeV and another at 1.5 GeV 
are seen in the sideband-subtracted mass spectrum. 
There is the hint of a possible peak (or fluctuation) near 1.75 GeV, 
but it is not statistically significant and will be investigated 
in future higher-statistics data sets acquired with CLAS12.

\subsubsection{Momentum Transfer Cut \label{sub:t-cuts} }

In the invariant mass spectrum in Fig. \ref{fig:Mkkbs} the resonance
of interest is the one at 1.50 GeV. In order to further investigate
it, cuts on the momentum transfer, $t$, were applied,
\begin{equation}
t = ( P_{beam} - P_{K_S^0}^1 - P_{K_S^0}^2 )^{2} \ ,
\end{equation}
where $P_{K_S^0}^{1,2}$ are the 4-momenta of the two $K_S^0$, each 
made from the 4-momenta of two charged pions.

\begin{figure*}
\begin{centering}
\includegraphics[scale=0.40]{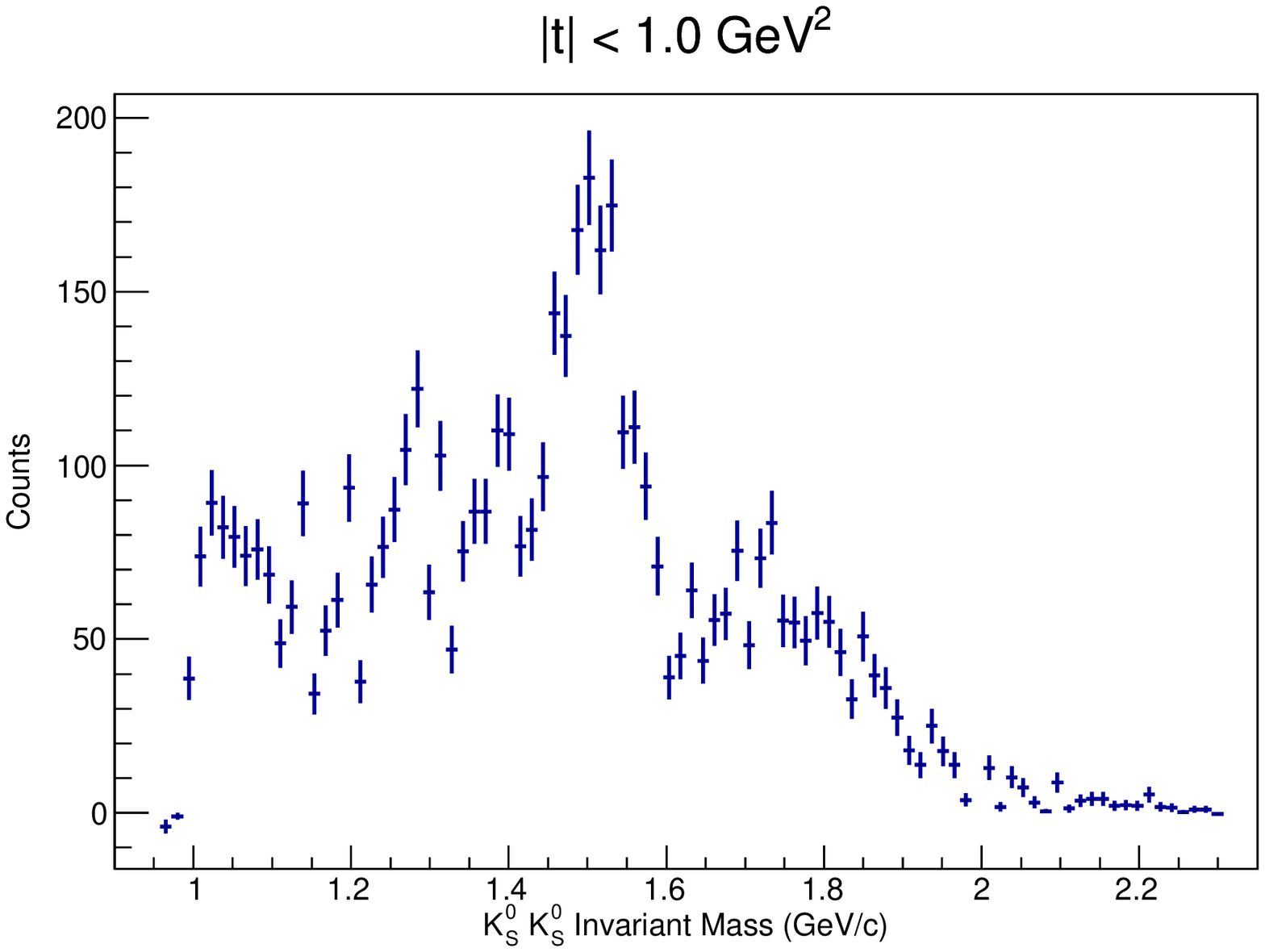} 
\includegraphics[scale=0.40]{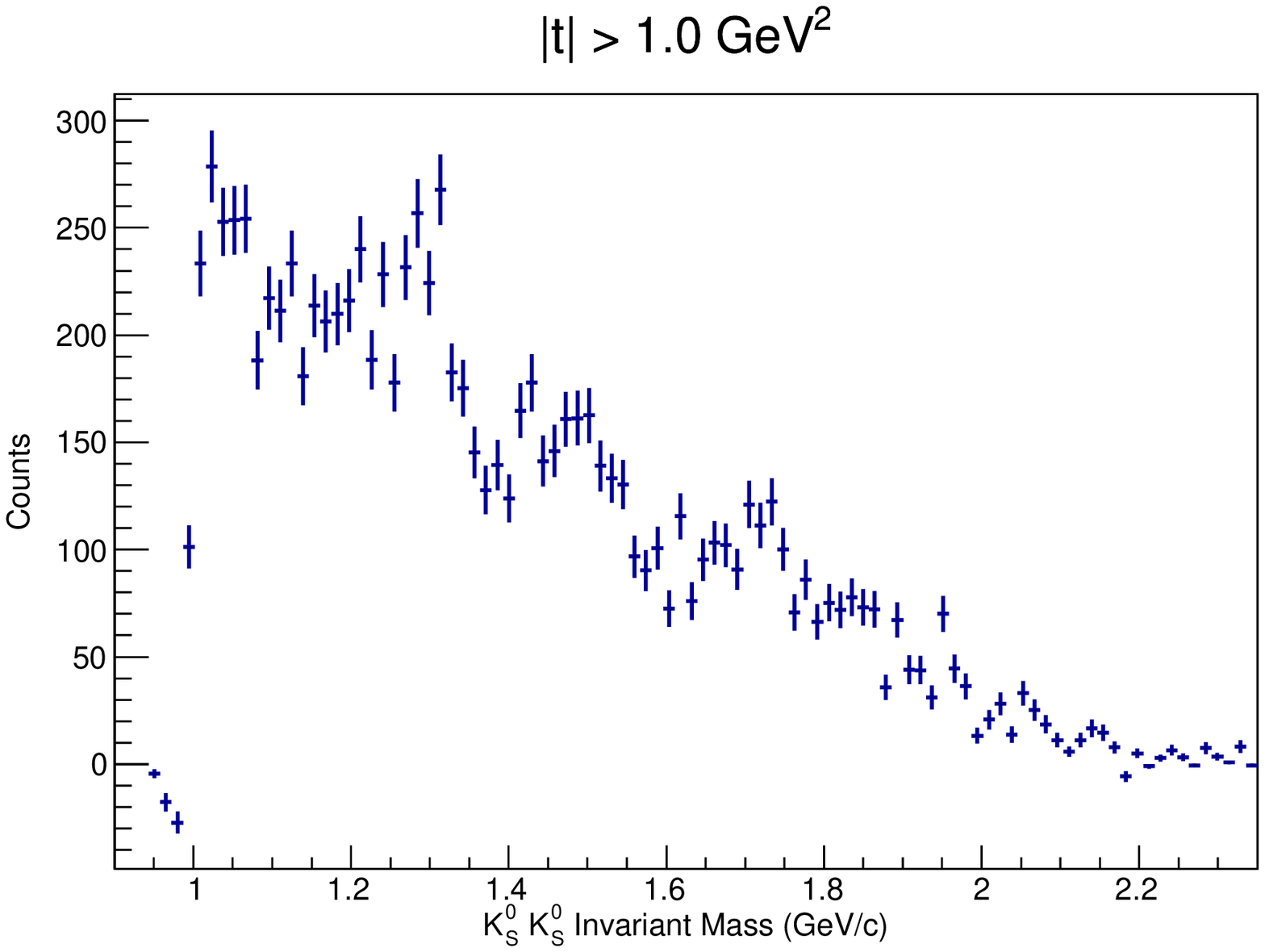} 
\par\end{centering}
\protect\caption{
Background subtracted plots for the $4\pi$ invariant mass for 
$|t|<1$ GeV$^{2}$ (left) and $|t|>1$ GeV$^{2}$ (right). 
\label{fig:t-cut} }
\end{figure*}

In Fig. \ref{fig:t-cut} (left), where the cut $|t|<1$ GeV$^{2}$ has been 
applied, the 1.50 GeV resonance is enhanced in the spectrum,
whereas it disappears for $|t|>1$ GeV$^{2}$, as shown in 
Fig. \ref{fig:t-cut} (right). 
If an $s$-channel production mechanism
was involved, we would have expected to see the peak over a wider 
range of $t$ (within the available phase space). 
The $t$-dependence of the peak at 1.50 GeV is consistent
with a meson exchange process 
(a $t$-channel diagram, Fig. \ref{fig:s-and-t-channel}).

The choice to cut at $|t|=1$ GeV$^2$ is somewhat arbitrary, but is a 
reasonable attempt to separate small and large momentum transfer.  
For example, if there is $\rho$-exchange in a $t$-channel diagram, 
this would contribute more significantly at $|t|<1$ GeV$^2$, where the 
momentum transfer is a better match with the mass of the $\rho$ meson.
Choosing a slightly different value for the cut on $|t|$ does not 
change our conclusions.

\begin{figure*}
\begin{centering}
\includegraphics[scale=0.40]{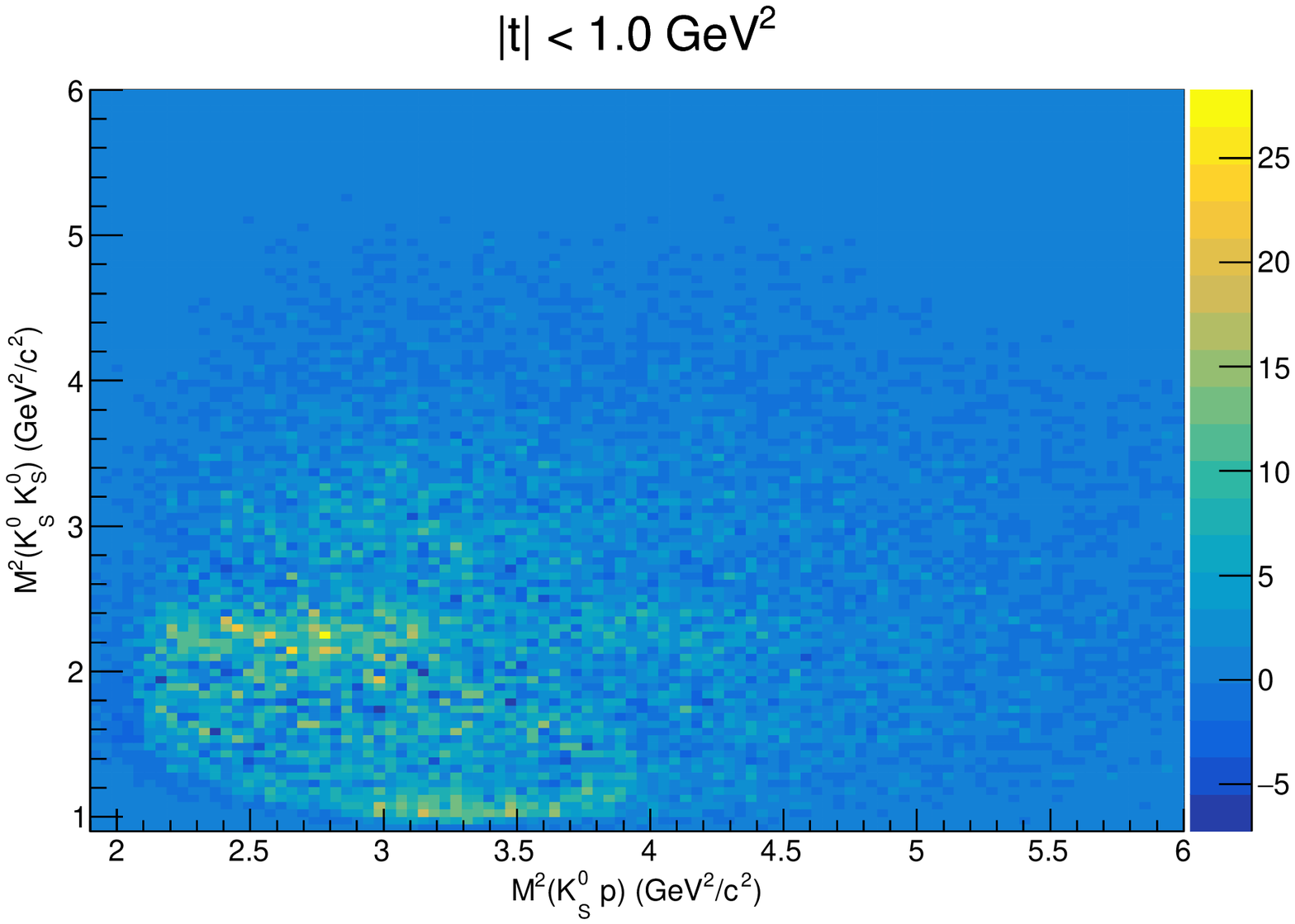}
\includegraphics[scale=0.40]{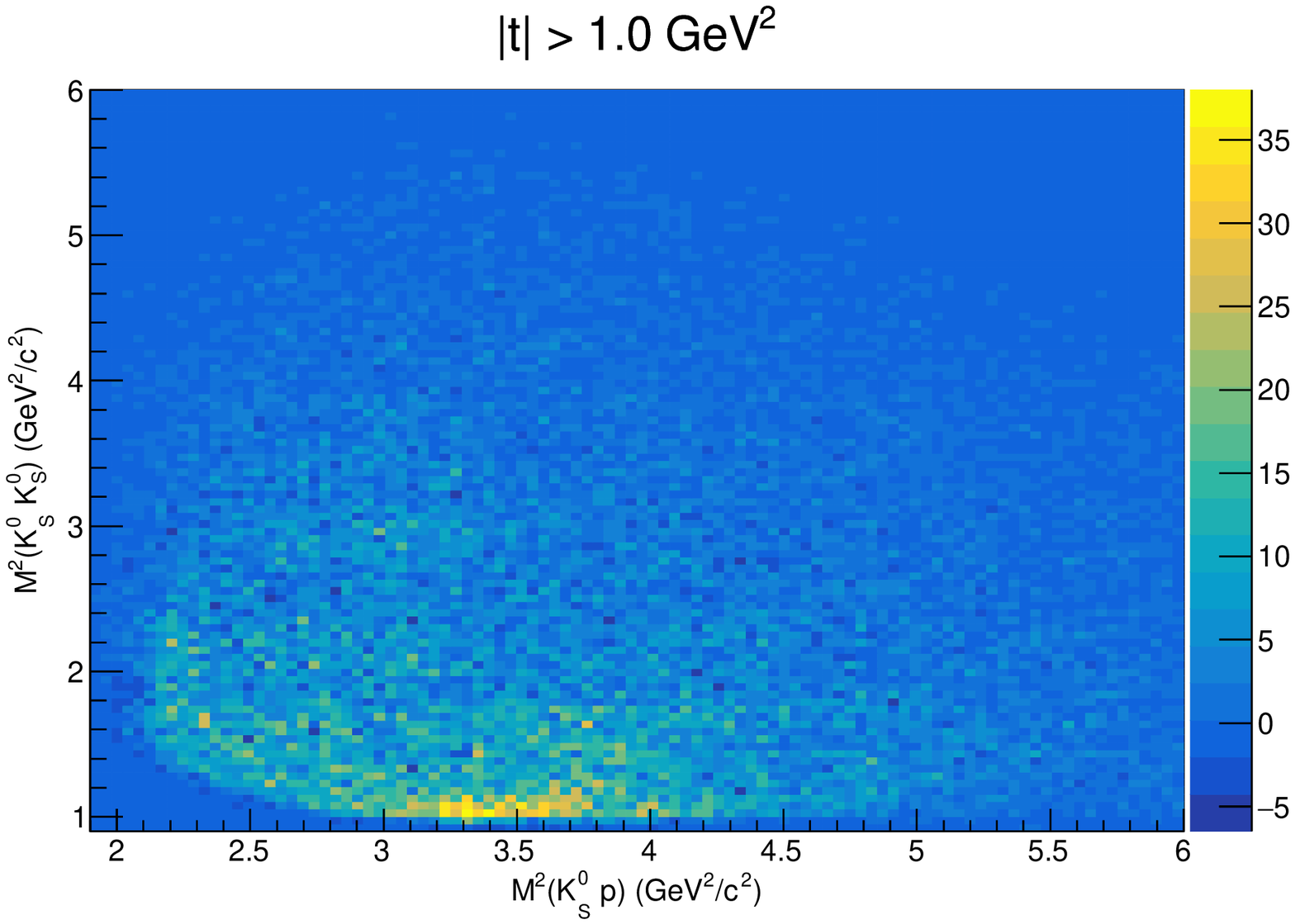}
\par\end{centering}
\protect\caption{
Dalitz plots of the three decay particles, the two kaons and the proton.
Structures making horizontal bands represent two-kaon resonances.
The intense region at the bottom, near $M^2(K_S^0,K_S^0)=1.0$ GeV$^2$, is 
due to the $f_0(980)$ decay.
\label{fig:Dalitz}}
\end{figure*}

\subsection{Dalitz Plots to Look for Baryon Resonances in Background}

To look for any possible background baryon resonances decaying into 
$K_S^0$ and $p$, Dalitz plots of $M^2(K_S^0,K_S^0)$ vs.~$M^2(K_S^0,p)$, 
where $M^2(X,Y)$ is the squared invariant mass of particles $X$ and $Y$,
are plotted in Fig. \ref{fig:Dalitz} for $|t|<1$ and $|t|>1$~GeV$^2$. 
These plots include the application of all cuts from 
Table \ref{tab:The-basic-cuts}, as well as the momentum transfer cut, 
and hence are the events remaining in the signal region after sideband 
subtraction has been done. 
The sideband subtraction was done on a bin by bin basis.

In Fig. \ref{fig:Dalitz}, the only structures seen are the horizontal 
bands, which represent resonances of two $K_S^0$ mesons. 
In the $|t|<1$ GeV$^{2}$ plot, the horizontal band at
2.25 GeV$^{2}$ is at the squared mass of the 1.50 GeV peak.
Also, the influence of the $f_{0}(980)$ is seen in the $|t|>1$
GeV$^{2}$ plot as a horizontal band near 1 GeV$^{2}$. 
The lack of any vertical structure indicates that no baryon 
resonances survive in the sideband-subtracted signal region. 
Even looking at the Dalitz plots before background subtraction 
(not shown), no clear vertical structures corresponding to baryon 
resonances can be seen. 
This is likely due to the cut on $E_{\gamma^{min}}>2.7$ 
GeV, which puts the center-of-mass energy, $W$, above the region 
where any narrow hyperon resonances could be seen.

\section{Simulations}

\subsection{Modeling the CLAS Detector}

In order to study the acceptance of CLAS, 
$\gamma p \rightarrow K_S^0 K_S^0 p$ events with 
$K_S^0 K_S^0 \rightarrow \pi^+ \pi^- \pi^+ \pi^-$
were generated isotropically with no dependence on $t$ 
for the purpose of comparing the data to pure phase space. 
The incident electron energy was set at 5.7 GeV, which translated 
into tagged bremsstrahlung photon beam energies of 1.5 GeV to 5.45 GeV. 
The target was positioned in the simulations exactly as in the g12 run.

These generated events were then passed to a program called GSIM
(Geant SIMulation) that models the CLAS detector using GEANT3
libraries, and digitizes the information. After being processed through
GSIM, the events are passed through a post processor,
which accounted for the condition of the CLAS detector during the g12
experimental run period. Using the g12 run conditions, 
the post processor removed hits that came from non-functioning 
parts of the detector and smeared values of measurements depending on the
resolution of the corresponding detector element during the g12 run
period. These processed events are then fed into the standard 
CLAS reconstruction software. Details of the reconstruction
process are given in Ref. \cite{key-12}.

\begin{figure*}
\includegraphics[scale=0.40]{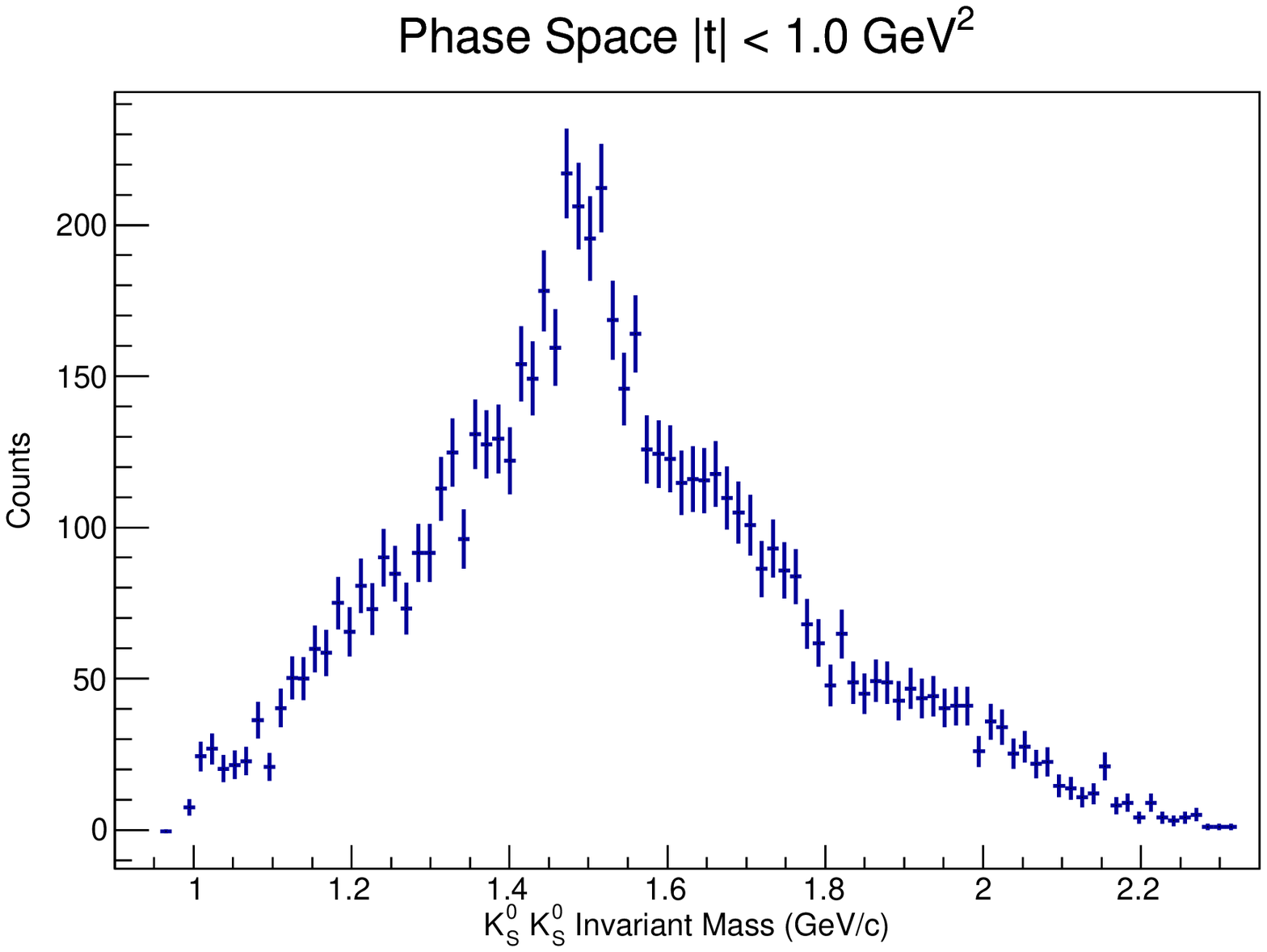}
\includegraphics[scale=0.40]{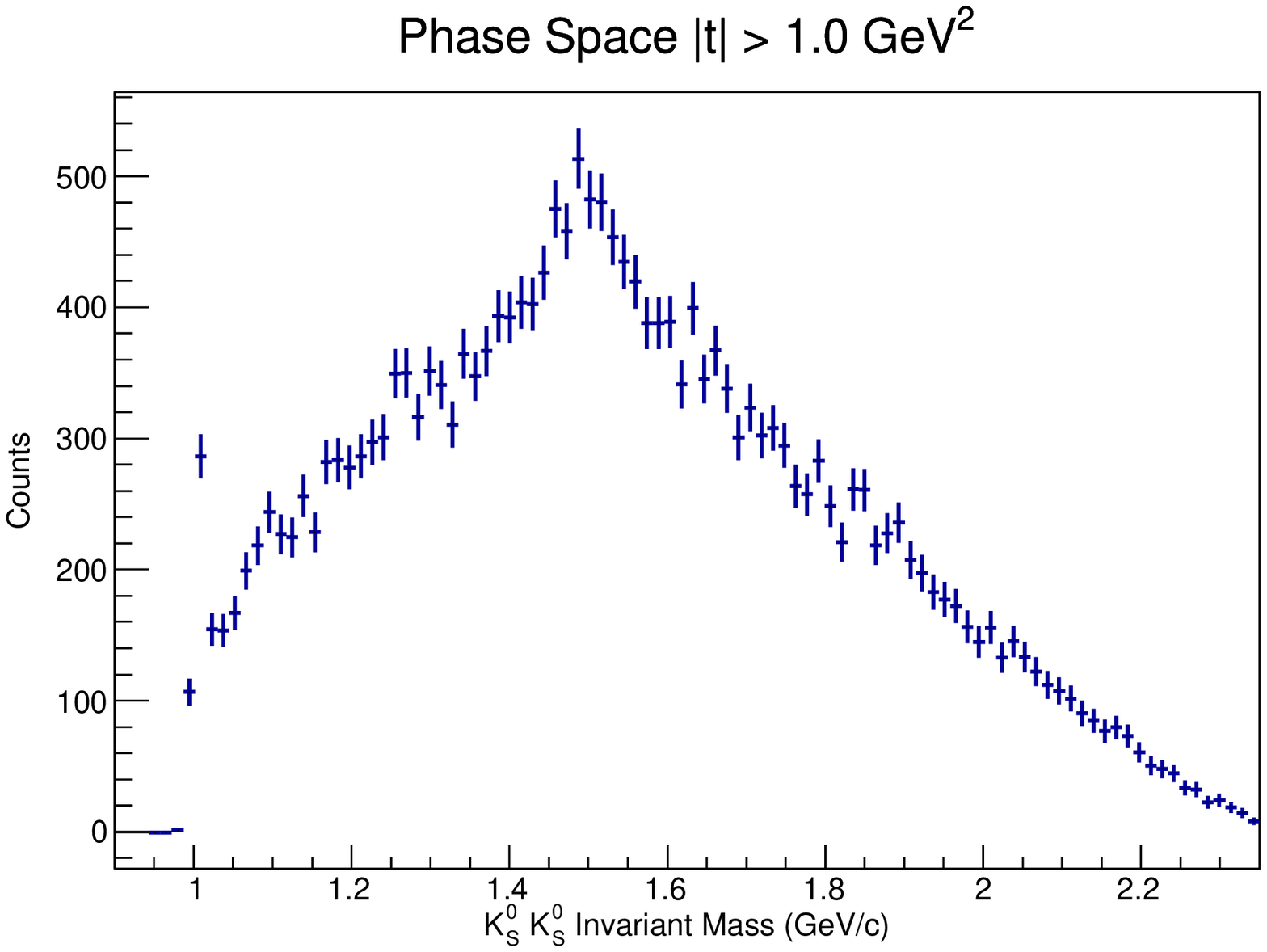}
\protect\caption{
Invariant mass spectrum of $4\pi$ phase space plus simulated 
$f_{0}(1500)$ and $f_{0}(980)$ mesons decaying to two $K_S^0$. 
Momentum transfer cuts $|t|<1$ GeV$^{2}$ (left) and 
$|t|>1$ GeV$^{2}$ (right) are shown.
\label{fig:t-cut+phasespace+f0} }
\end{figure*}

\begin{figure*}
\begin{centering}
\includegraphics[scale=0.40]{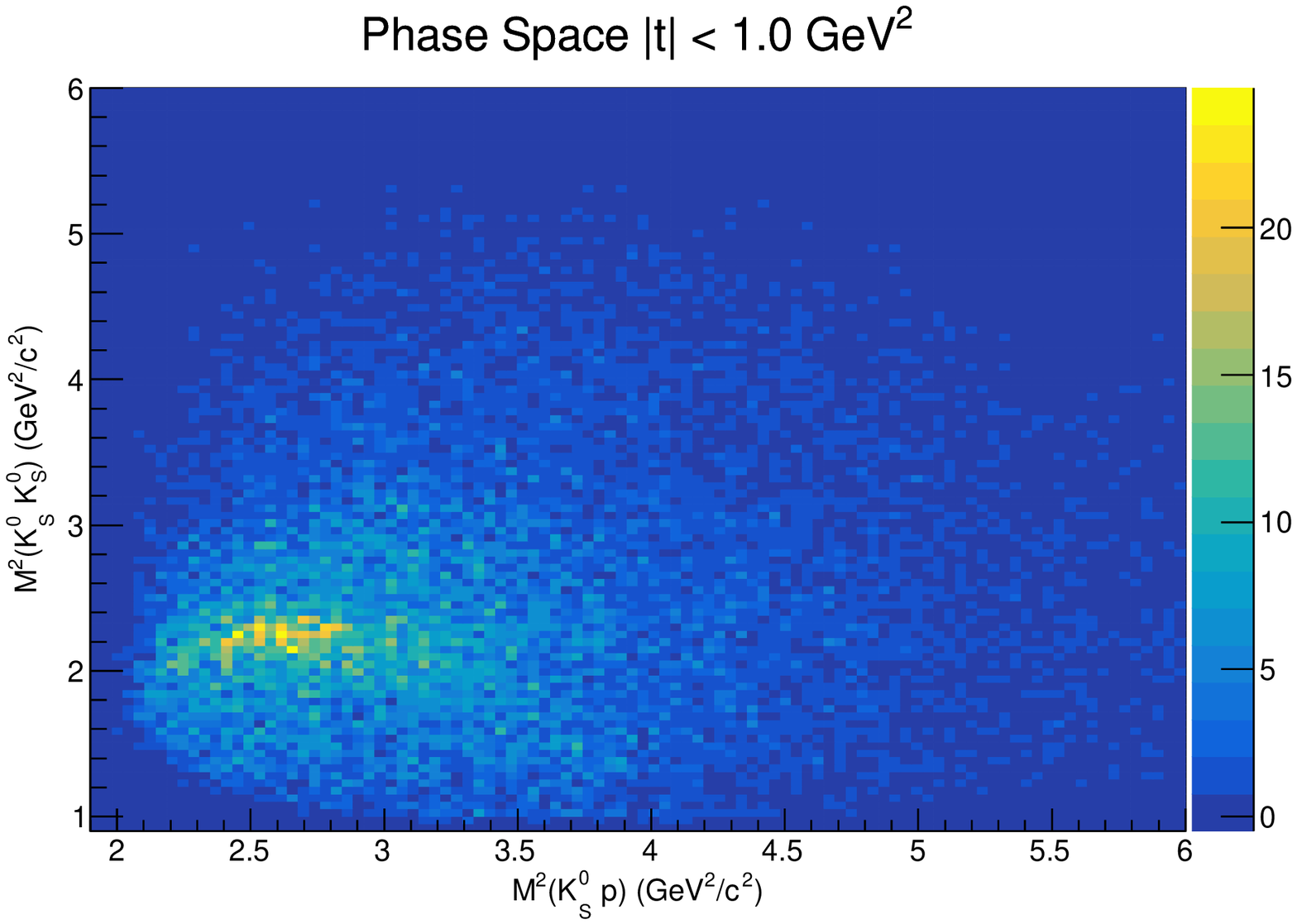}
\includegraphics[scale=0.40]{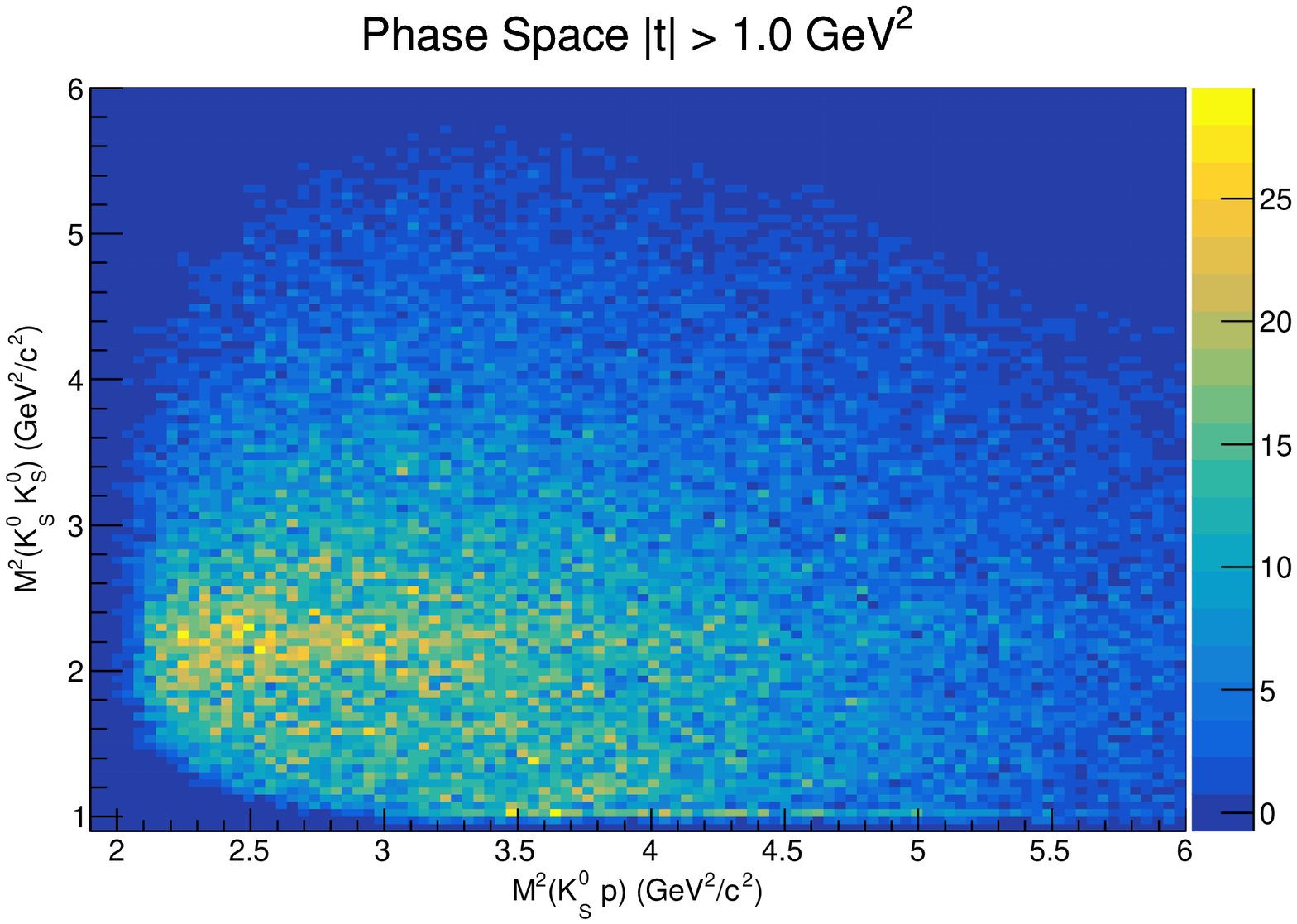}
\par\end{centering}
\protect\caption{
Dalitz plots of simulated events for $M^2(K_S^0,K_S^0)$ vs.~$M^2(K_S^0,p)$ 
for $|t|<1$ GeV$^{2}$ (left) and $|t|>1$ GeV$^{2}$ (right) for 
generated phase space plus $f_{0}(980)$ and $f_{0}(1500)$ mesons. 
\label{fig:dal+phasespace+f0} }
\end{figure*}

\subsection{Simulations: Phase Space, $f_0(980)$ and $f_0(1500)$ }

The Monte Carlo events that passed through the reconstruction software 
were then fed through the same analysis code as for the real data. 
The events remaining after this are called \emph{accepted} events. 
The upper tail of the $f_{0}(980)$ can decay into two kaons and can 
be distinctly seen in Fig. \ref{fig:Dalitz}, in addition to the 
horizontal band due to the 1.5 GeV resonance. 
Separate simulations were carried out for 
$\gamma p\rightarrow f_{0}(980) p$ and 
$\gamma p\rightarrow f_{0}(1500) p$, and these were then added
to the phase space Monte Carlo (MC) events. 
Cuts were made to divide the simulated four-pion invariant mass spectrum 
into two sets, one with $|t|<1$~GeV$^{2}$ and the other with 
$|t|>1$~GeV$^{2}$, as shown in Fig. \ref{fig:t-cut+phasespace+f0}. 

The simulated peak at 1.50 GeV, from the $f_0(1500)$ decay, 
is present to a larger extent for $|t|>1$ than for $|t|<1$ GeV$^{2}$. 
The increased number of counts of the 1.50 GeV peak in the simulations 
at higher momentum transfer $|t|$ is expected kinematically, 
due to the increase of the available phase space.  
This is reiterated in the Dalitz plots shown in 
Fig. \ref{fig:dal+phasespace+f0}. 
The comparison of the real data with the phase-space MC simulations 
reinforces the idea that the physical process associated with 
the production of the peak at 1.50 GeV is from a $t$-channel process.

\begin{figure*}
\noindent \begin{raggedright}
\includegraphics[scale=0.80]{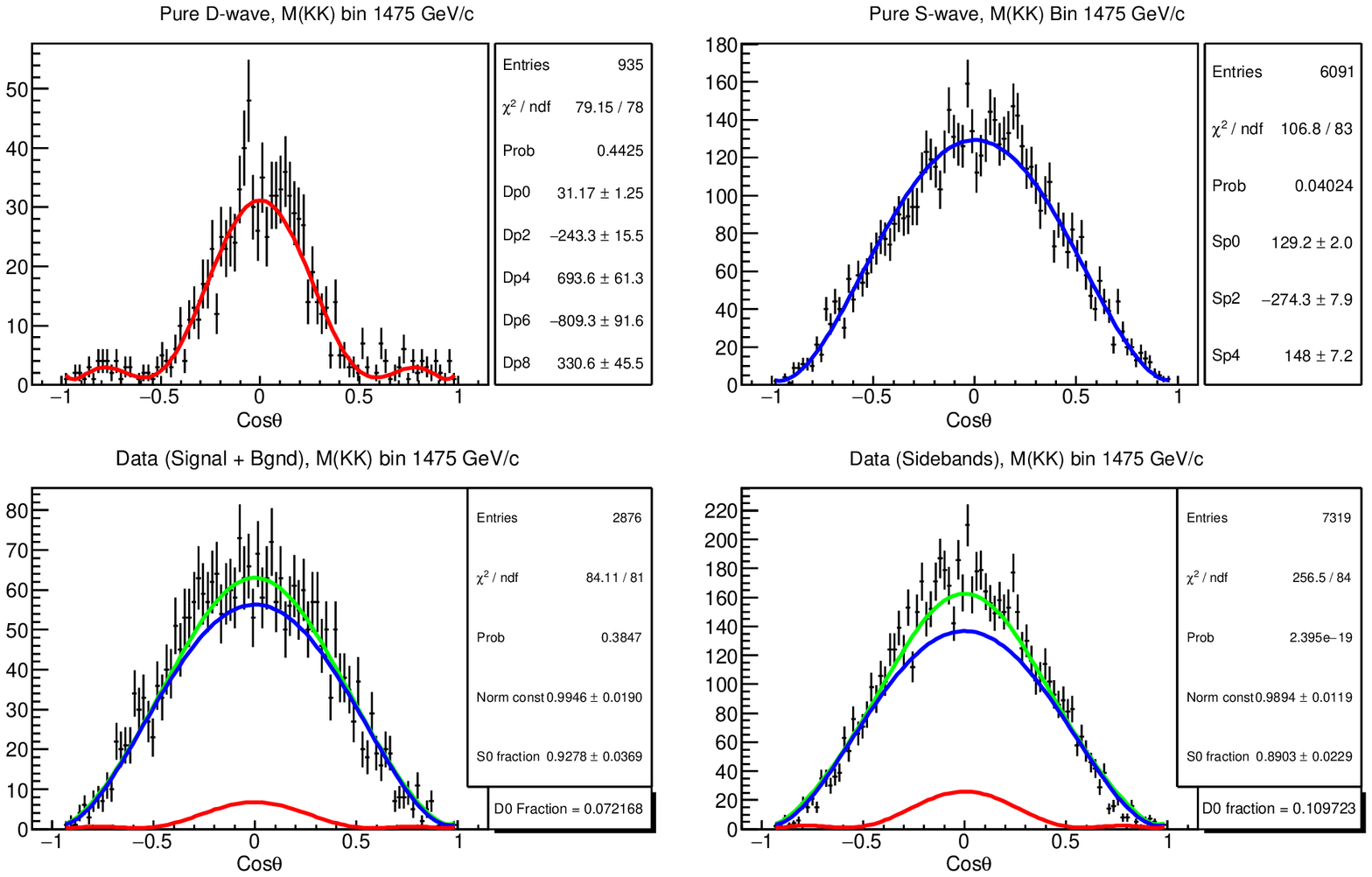}
\par\end{raggedright}
\protect\caption{
Fits to $\cos\theta_{c.m.}$ distributions in the Gottfied-Jackson frame for 
simulated pure $S$ wave and pure $D$ wave (top) 
and data from the S+B and sideband regions (bottom) 
for bin 1450-1500 MeV of the $K_S^0 K_S^0$ mass. 
The fit curves are: $S$ wave (red online), $D$ wave (blue online) 
and Total (green online).
The parameters of the fits are shown to the right of the plots.
\label{fig:Fits-to-1475} }
\end{figure*}

\begin{figure*}
\noindent \begin{raggedright}
\includegraphics[scale=0.80]{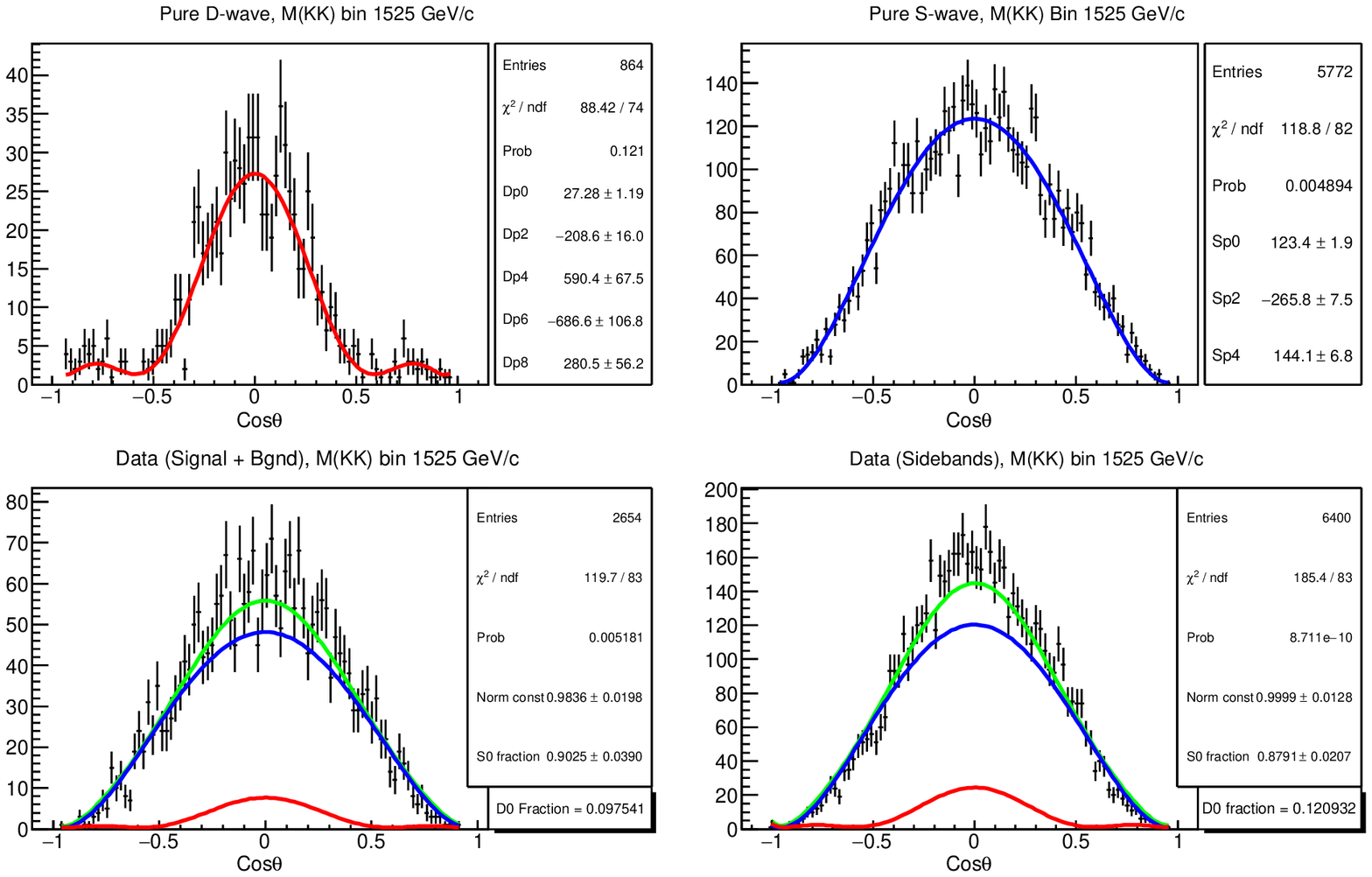}
\par\end{raggedright}
\protect\caption{
Same as in Fig. \ref{fig:Fits-to-1475}, 
but for bin 1500-1550 MeV of the $K_S^0 K_S^0$ mass.}
\end{figure*}

\begin{figure*}
\noindent \begin{raggedright}
\includegraphics[scale=0.80]{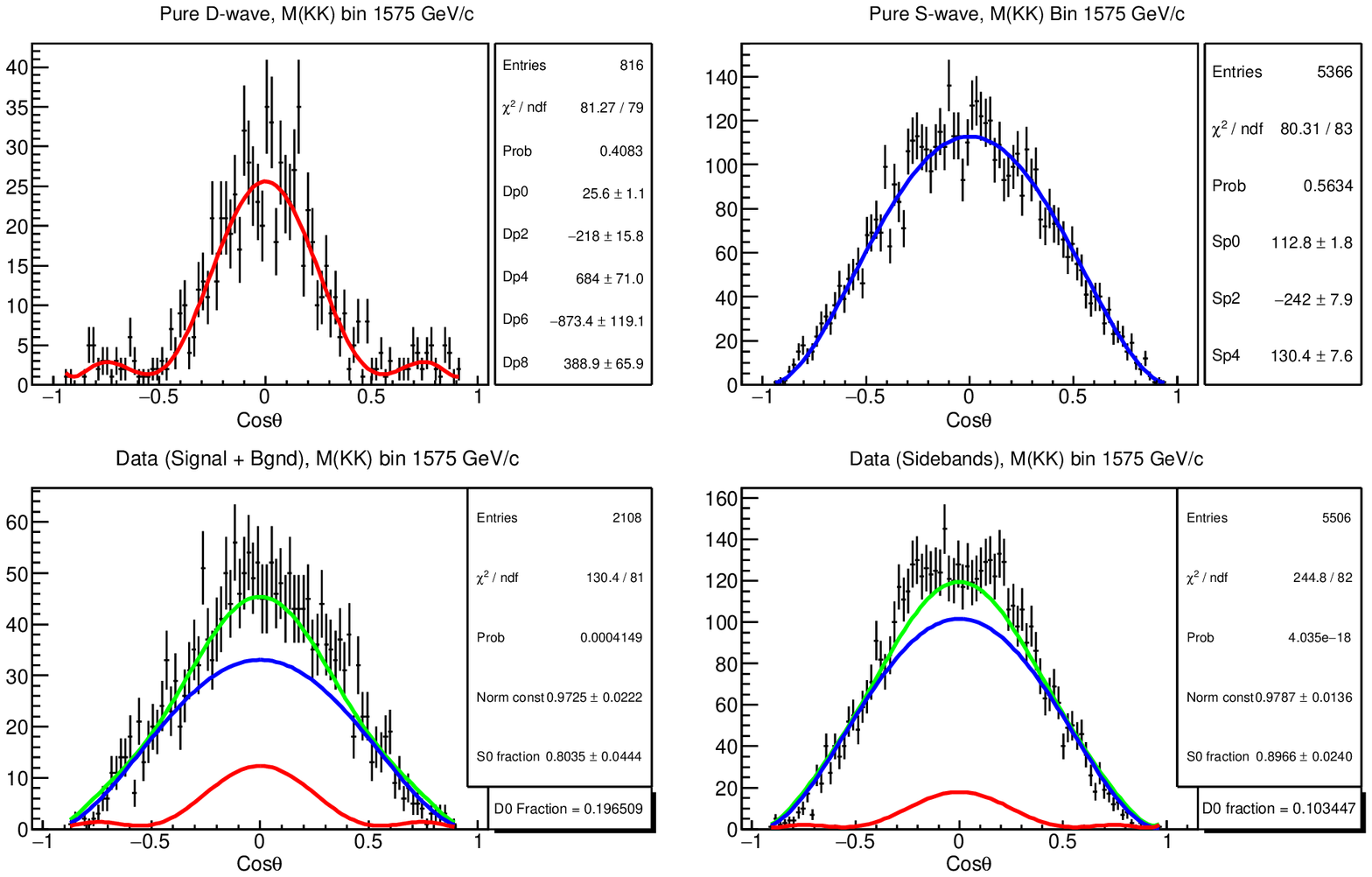}
\par\end{raggedright}
\protect\caption{
Same as in Fig. \ref{fig:Fits-to-1475}, 
but for bin 1550-1600 MeV of the $K_S^0 K_S^0$ mass.
\label{fig:Fits-to-1575} }
\end{figure*}

\section{Results }

In this section, the polar angular distributions of the data and MC
are examined in order to extricate the spin contributions from $J=0$
and $J=2$.

The data and Monte Carlo events were binned in 50 MeV intervals of 
the two-$K_S^0$ invariant mass. The low statistics of the data do 
not allow for further binning in $t$ or $E_{\gamma}$ while still 
providing sufficiently accurate angular distributions. 
Hence in our analysis of the angular distributions, we examine both the 
signal + background (S+B) and the sideband background regions, 
drawing our conclusions based on the comparison of these two regions. 
The evaluation of the angular distributions of these spectra
begins with the generation of simulated pure $S$ and $D$ waves. 
The phase space distribution behaves like an $S$ wave, so these angular 
distributions can be obtained by the MC generating phase space. 
A pure $D$ wave was generated in the Gottfried-Jackson frame
and run through the reconstruction software.

Figures \ref{fig:Fits-to-1475}-\ref{fig:Fits-to-1575} portray the
polar angle distributions (and the fits to it) for the $S$ wave, $D$ wave, 
data (S+B), and data (sideband) regions for mass bins ranging from 
1450-1600 MeV. 
The polar angle distributions of the simulated pure $S$ and $D$ waves 
were fit with polynomials including only even orders of 
$\cos\theta_{c.m.}$ to preserve symmetry.  In order to compare these with 
data, plots of the polar angle in the Gottfried-Jackson frame 
(after passing through the detector simulations) were normalized 
such that they have nearly the same number of events as the data.

The distributions of the S+B and sideband regions were fit using
the functional shapes extracted from fitting the pure $S$ and $D$ wave 
angular distributions. The formula used is:
\begin{equation}
Total = N \left( f \cdot (S_{wave}) 
      + (1 - f) \cdot (D_{wave}) \right) \ ,
\end{equation}
where $N$ is a normalization constant and $f$ is the fraction of 
$S$ wave strength. 
This fitting {\it assumes} no interference of $S$ and $D$ waves;
in reality we know that interference occurs, but the 
detector acceptance for CLAS is not uniform and this creates 
a bi-modal ambiguity in attempts to separate the $S$ and $D$ waves 
when interference is included in the fits.
The above equation provides the most practical indication,
within the limitations of the CLAS detector acceptance, of the 
$S$ and $D$ wave fractions from each mass bin.
The lowest (red online) curve denotes the function describing the D-wave, 
the middle (blue online) one denotes the $S$ wave and 
the top (green online) curve is the total fit.

\begin{table*}
\protect\caption{Fraction of $S$ wave from fits to the S+B and 
sideband regions.
\label{tab:S-fractions}}
\begin{centering}
\begin{tabular}{|c|c|c|}
\hline 
Mass Bin & $S$ wave fraction & $S$ wave fraction \\
 (MeV) & (S+B region) & (Sidebands) \\
\hline 
\hline 
1000-1050 & 1.000 $\pm$ 0.045 & 1.000 $\pm$ 0.031 \\
\hline 
1050-1100 & 1.000 $\pm$ 0.031 & 1.000 $\pm$ 0.029 \\
\hline 
1100-1150 & 0.973 $\pm$ 0.025 & 0.982 $\pm$ 0.018 \\
\hline 
1150-1200 & 1.000 $\pm$ 0.023 & 1.000 $\pm$ 0.015 \\
\hline 
1200-1250 & 1.000 $\pm$ 0.022 & 1.000 $\pm$ 0.011 \\
\hline 
1250-1300 & 1.000 $\pm$ 0.013 & 1.000 $\pm$ 0.063 \\
\hline 
1300-1350 & 1.000 $\pm$ 0.020 & 1.000 $\pm$ 0.011 \\
\hline 
1350-1400 & 1.000 $\pm$ 0.028 & 1.000 $\pm$ 0.026 \\
\hline 
1400-1450 & 1.000 $\pm$ 0.025 & 0.922 $\pm$ 0.019 \\
\hline 
1450-1500 & 0.928 $\pm$ 0.037 & 0.890 $\pm$ 0.023 \\
\hline 
1500-1550 & 0.903 $\pm$ 0.039 & 0.879 $\pm$ 0.021 \\
\hline 
1550-1600 & 0.803 $\pm$ 0.044 & 0.897 $\pm$ 0.024 \\
\hline 
1600-1650 & 0.791 $\pm$ 0.056 & 0.883 $\pm$ 0.032 \\
\hline 
1650-1700 & 0.762 $\pm$ 0.052 & 0.910 $\pm$ 0.031 \\
\hline 
1700-1750 & 0.660 $\pm$ 0.053 & 0.902 $\pm$ 0.033 \\
\hline 
1750-1800 & 0.690 $\pm$ 0.071 & 0.941 $\pm$ 0.041 \\
\hline 
1800-1850 & 0.845 $\pm$ 0.086 & 0.994 $\pm$ 0.096 \\
\hline 
\end{tabular}
\par\end{centering}
\end{table*}

The values of the fraction of $S$ wave present in the two regions, based
on the fits, are tabulated in Table \ref{tab:S-fractions}. 
Both the S+B and sideband regions show mostly the shape of a pure $S$ wave
up to the 1400 MeV mass bin. 
In the mass regions 1450-1500 and 1500-1550 MeV, 
both of which include the peak of interest, the S+B regions have 
slightly smaller $D$ wave fractions than the corresponding sideband regions, 
which suggests that the signal is mostly $S$ wave in these mass bins. 
The higher mass bins (as for the 1550-1600 MeV bin in Fig. 
\ref{fig:Fits-to-1575}) 
involve more $D$ wave shape in the S+B region 
than in the sidebands, which implies some amount of resonance 
contributions there. 
Since there are no well-known $J^{PC}=2^{++}$ resonances in the higher 
mass bins, we do not speculate as to the possible influence of resonances 
contributions there. 
The implications of the peak at 1.50 GeV will be discussed next.

\section{Summary and Discussion}

In this analysis, the reaction $\gamma p\to pX \to pK_S^0 K_S^0$
was investigated using data from the g12 experiment at Jefferson Lab. 
This represents the first high statistics data for photoproduction of 
scalar mesons with masses above 1 GeV from the CLAS detector. 
Four charged pions were detected and missing mass was used to 
identify an exclusive final state.
Combinations of $\pi^+ \pi^-$ pairs clearly show correlations from the 
decay of two $K_S^0$ over a nearly flat four-pion background.
The two identical $K_S^0$ decay requires the parent meson to have a 
definite state of $CP=++$. 

The sideband-subtraction method was employed to obtain the $K_S^0 K_S^0$ 
(or four-pion) invariant mass spectrum, which shows peaks centered 
at 1.28 GeV and 1.50 GeV, with some background still present. 
The physics associated with this background is unknown,
and examination of Dalitz plots do not show significant background 
from any narrow hyperon resonances that could reflect into the 
invariant mass spectrum of the two $K_S^0$.

At first glance, the resonance at 1.28 GeV could easily be mistaken
for the $f_{2}(1270)$. However, the width of the observed peak is
much narrower than the average PDG listed width of the $f_{2}(1270)$,
so it is not clear if this bump represents a meson resonance 
or something else (such as a cusp effect).
The resonance at 1.50 GeV is distinctly seen at low momentum transfer,
but disappears above $|t|>$1 GeV$^2$, consistent with production via a
$t$-channel process. 

The low acceptance at forward and backward angles with CLAS for this 
final state prevents us from performing a full partial wave analysis. 
In light of this, to check for contributions from the lowest order 
symmetric waves, the angular distribution in the Gottfried-Jackson 
frame of the $K_S^0 K_S^0$ decay was compared with that of 
simulated pure $S$ and $D$ waves. 
Both S+B (signal + background) and sideband regions were separately 
fit to the decay shape extracted from $S$ and $D$ waves for each 
$K_S^0 K_S^0$ mass bin, and differences between the two gave an 
indication of which partial wave dominates the signal at that mass. 

The lower mass bins, from 1000 MeV to 1400 MeV, have almost 100\% 
$S$ wave contribution. For the 1450-1500 and 1500-1550 MeV bins, 
where the $f_0(1500)$ and $f_2'(1525)$ mesons are expected to contribute,
the S+B and sideband regions have similar contributions from 
$S$ and $D$ waves with slightly larger $S$ wave fractions in the S+B 
region, suggesting that the signal in this mass range is mostly $S$ wave. 
However, the assumption of no interference used in the fits (which 
is a necessary condition due to holes in the CLAS accceptance) 
prevents a firm conclusion on the $S$ or $D$ wave nature of the 
peak at 1500 MeV.
For bins above 1550 MeV, the $D$ wave fraction in the S+B region is 
greater than that in the sidebands, implying some $D$ wave in the signal.

In conclusion, fits to the angular distributions of the data 
suggest that most of the $K_S^0 K_S^0$ decay in the 1450-1550 MeV 
mass region is $S$ wave. In addition, the mass and width of the peak
at 1500 MeV is consistent with that of the $f_{0}(1500)$. For these
reasons, we propose that the observed resonance at 1.50 GeV in Figs.
\ref{fig:Mkkbs} and \ref{fig:t-cut} is most likely from the $S$ wave 
$f_{0}(1500) \to K_S^0 K_S^0$. Since this resonance is seen mostly 
at low momentum transfer ($|t|<1$ GeV$^2$), consistent with $t$-channel 
meson production, we speculate that the glueball content of this 
resonance is not large. If confirmed, this result would suggest 
that the $f_0(1710)$ is the more likely candidate to have a high 
overlap with the lowest glueball state, consistent with recent 
theoretical indications \cite{key-6}. 

The $f'_{2}(1525)$ has a mass of 1525 MeV and
a width of 73 MeV, and hence there is a possibility of it contributing
to this mass region in our data. Although the results from the decay 
angular fits are consistent with the presence of the $f_{0}(1500)$, 
a contribution from the $f'_{2}(1525)$ cannot be ruled out.

This is the first time that this final state has been analyzed in
photoproduction and hence it contributes new information to the world
data on scalar mesons. Future experiments with the luminosities now 
available at CLAS12 \cite{key-16} and GlueX at Jefferson Lab might 
afford better statistics and better acceptance for a more definitive 
study of this final state.

\begin{acknowledgments}
The authors gratefully acknowledge the work of Jefferson Lab staff 
in the Accelerator and Physics Divisions. 
This work was supported by: the United Kingdom's Science
and Technology Facilities Council (STFC); 
the Chilean Comisi\`on Nacional de Investigaci\`on Cient\`ifica y 
Tecnol\`ogica (CONICYT); 
the Italian Istituto Nazionale di Fisica Nucleare; 
the French Centre National de la Recherche Scientifique; 
the French Commissariat \`a l'Energie Atomique;
the U.S. National Science Foundation; 
and the National Research Foundation of Korea. 
Jefferson Science Associates, LLC, operates the Thomas Jefferson 
National Accelerator Facility for the the U.S. Department of Energy 
under Contract No. DE-AC05-06OR23177. 
\end{acknowledgments}

\end{document}

%% file: auth_revtex.tex
\newcommand*{\OHIOU}{Ohio University, Athens, Ohio  45701}
\newcommand*{\OHIOUindex}{1}
\affiliation{\OHIOU}
\newcommand*{\VIRGINIA}{University of Virginia, Charlottesville, Virginia 22901}
\newcommand*{\VIRGINIAindex}{2}
\affiliation{\VIRGINIA}
\newcommand*{\Juelich}{Institute fur Kernphysik (Juelich), Juelich, Germany}
\newcommand*{\Juelichindex}{3}
\affiliation{\Juelich}
\newcommand*{\TEMPLE}{Temple University,  Philadelphia, PA 19122 }
\newcommand*{\TEMPLEindex}{4}
\affiliation{\TEMPLE}
\newcommand*{\CNU}{Christopher Newport University, Newport News, Virginia 23606}
\newcommand*{\CNUindex}{5}
\affiliation{\CNU}
\newcommand*{\ANL}{Argonne National Laboratory, Argonne, Illinois 60439}
\newcommand*{\ANLindex}{6}
\affiliation{\ANL}
\newcommand*{\ASU}{Arizona State University, Tempe, Arizona 85287-1504}
\newcommand*{\ASUindex}{7}
\affiliation{\ASU}
\newcommand*{\CSUDH}{California State University, Dominguez Hills, Carson, CA 90747}
\newcommand*{\CSUDHindex}{8}
\affiliation{\CSUDH}
\newcommand*{\CMU}{Carnegie Mellon University, Pittsburgh, Pennsylvania 15213}
\newcommand*{\CMUindex}{9}
\affiliation{\CMU}
\newcommand*{\CUA}{Catholic University of America, Washington, D.C. 20064}
\newcommand*{\CUAindex}{10}
\affiliation{\CUA}
\newcommand*{\SACLAY}{IRFU, CEA, Universit'e Paris-Saclay, F-91191 Gif-sur-Yvette, France}
\newcommand*{\SACLAYindex}{11}
\affiliation{\SACLAY}
\newcommand*{\UCONN}{University of Connecticut, Storrs, Connecticut 06269}
\newcommand*{\UCONNindex}{12}
\affiliation{\UCONN}
\newcommand*{\FU}{Fairfield University, Fairfield CT 06824}
\newcommand*{\FUindex}{13}
\affiliation{\FU}
\newcommand*{\FERRARAU}{Universita' di Ferrara , 44121 Ferrara, Italy}
\newcommand*{\FERRARAUindex}{14}
\affiliation{\FERRARAU}
\newcommand*{\FIU}{Florida International University, Miami, Florida 33199}
\newcommand*{\FIUindex}{15}
\affiliation{\FIU}
\newcommand*{\FSU}{Florida State University, Tallahassee, Florida 32306}
\newcommand*{\FSUindex}{16}
\affiliation{\FSU}
\newcommand*{\GWUI}{The George Washington University, Washington, DC 20052}
\newcommand*{\GWUIindex}{17}
\affiliation{\GWUI}
\newcommand*{\ISU}{Idaho State University, Pocatello, Idaho 83209}
\newcommand*{\ISUindex}{18}
\affiliation{\ISU}
\newcommand*{\INFNFE}{INFN, Sezione di Ferrara, 44100 Ferrara, Italy}
\newcommand*{\INFNFEindex}{19}
\affiliation{\INFNFE}
\newcommand*{\INFNFR}{INFN, Laboratori Nazionali di Frascati, 00044 Frascati, Italy}
\newcommand*{\INFNFRindex}{20}
\affiliation{\INFNFR}
\newcommand*{\INFNGE}{INFN, Sezione di Genova, 16146 Genova, Italy}
\newcommand*{\INFNGEindex}{21}
\affiliation{\INFNGE}
\newcommand*{\INFNRO}{INFN, Sezione di Roma Tor Vergata, 00133 Rome, Italy}
\newcommand*{\INFNROindex}{22}
\affiliation{\INFNRO}
\newcommand*{\INFNTUR}{INFN, Sezione di Torino, 10125 Torino, Italy}
\newcommand*{\INFNTURindex}{23}
\affiliation{\INFNTUR}
\newcommand*{\ORSAY}{Institut de Physique Nucl\'eaire, CNRS/IN2P3 and Universit\'e Paris Sud, Orsay, France}
\newcommand*{\ORSAYindex}{24}
\affiliation{\ORSAY}
\newcommand*{\ITEP}{Institute of Theoretical and Experimental Physics, Moscow, 117259, Russia}
\newcommand*{\ITEPindex}{25}
\affiliation{\ITEP}
\newcommand*{\JMU}{James Madison University, Harrisonburg, Virginia 22807}
\newcommand*{\JMUindex}{26}
\affiliation{\JMU}
\newcommand*{\KNU}{Kyungpook National University, Daegu 41566, Republic of Korea}
\newcommand*{\KNUindex}{27}
\affiliation{\KNU}
\newcommand*{\MISS}{Mississippi State University, Mississippi State, MS 39762-5167}
\newcommand*{\MISSindex}{28}
\affiliation{\MISS}
\newcommand*{\UNH}{University of New Hampshire, Durham, New Hampshire 03824-3568}
\newcommand*{\UNHindex}{29}
\affiliation{\UNH}
\newcommand*{\NSU}{Norfolk State University, Norfolk, Virginia 23504}
\newcommand*{\NSUindex}{30}
\affiliation{\NSU}
\newcommand*{\ODU}{Old Dominion University, Norfolk, Virginia 23529}
\newcommand*{\ODUindex}{31}
\affiliation{\ODU}
\newcommand*{\RPI}{Rensselaer Polytechnic Institute, Troy, New York 12180-3590}
\newcommand*{\RPIindex}{32}
\affiliation{\RPI}
\newcommand*{\URICH}{University of Richmond, Richmond, Virginia 23173}
\newcommand*{\URICHindex}{33}
\affiliation{\URICH}
\newcommand*{\ROMAII}{Universita' di Roma Tor Vergata, 00133 Rome Italy}
\newcommand*{\ROMAIIindex}{34}
\affiliation{\ROMAII}
\newcommand*{\MSU}{Skobeltsyn Institute of Nuclear Physics, Lomonosov Moscow State University, 119234 Moscow, Russia}
\newcommand*{\MSUindex}{35}
\affiliation{\MSU}
\newcommand*{\SCAROLINA}{University of South Carolina, Columbia, South Carolina 29208}
\newcommand*{\SCAROLINAindex}{36}
\affiliation{\SCAROLINA}
\newcommand*{\JLAB}{Thomas Jefferson National Accelerator Facility, Newport News, Virginia 23606}
\newcommand*{\JLABindex}{37}
\affiliation{\JLAB}
\newcommand*{\UTFSM}{Universidad T\'{e}cnica Federico Santa Mar\'{i}a, Casilla 110-V Valpara\'{i}so, Chile}
\newcommand*{\UTFSMindex}{38}
\affiliation{\UTFSM}
\newcommand*{\EDINBURGH}{Edinburgh University, Edinburgh EH9 3JZ, United Kingdom}
\newcommand*{\EDINBURGHindex}{39}
\affiliation{\EDINBURGH}
\newcommand*{\GLASGOW}{University of Glasgow, Glasgow G12 8QQ, United Kingdom}
\newcommand*{\GLASGOWindex}{40}
\affiliation{\GLASGOW}
\newcommand*{\VT}{Virginia Tech, Blacksburg, Virginia   24061-0435}
\newcommand*{\VTindex}{41}
\affiliation{\VT}
\newcommand*{\WM}{College of William and Mary, Williamsburg, Virginia 23187-8795}
\newcommand*{\WMindex}{42}
\affiliation{\WM}
\newcommand*{\YEREVAN}{Yerevan Physics Institute, 375036 Yerevan, Armenia}
\newcommand*{\YEREVANindex}{43}
\affiliation{\YEREVAN}

 %%%%%%%%%%%%%%% END OF Latex Macros for institute addresses  %%%%%%%%%%%%%%%%%%%%%%%%% 

\author {S.~Chandavar} 
\affiliation{\OHIOU}
\author {J.T.~Goetz} 
\affiliation{\OHIOU}
\author {K.~Hicks} 
\affiliation{\OHIOU}
\author {D.~Keller} 
\affiliation{\OHIOU}
\affiliation{\VIRGINIA}
\author {M.C.~Kunkel} 
\affiliation{\Juelich}
\author {M.~Paolone} 
\affiliation{\TEMPLE}
\author {D.P.~Weygand} 
\affiliation{\CNU}
\author {K.P.~Adhikari} 
\affiliation{\MISS}
\affiliation{\ODU}
\author {S.~Adhikari} 
\affiliation{\FIU}
\author {Z.~Akbar} 
\affiliation{\FSU}
% ...

\collaboration{The CLAS Collaboration}
\noaffiliation